\documentstyle[12pt,draft]{IEEEtran}
\begin{document}
\input{psfig}

\title{Self-Control of Chaotic Dynamics using LTI Filters}

\author{Pabitra Mitra\thanks{Author is with the Institute for Robotics and Intelligent Systems, Bangalore, India.  EMail: pabitra@cair.res.in}}

\maketitle
\thispagestyle{plain}\pagestyle{plain}

\begin{abstract}
In this brief, an algorithm for controlling chaotic systems using small, continuous time perturbations is presented. Stabilisation is achieved by self controlling feedback using low order LTI filters. The algorithm alleviates the need of complex calculations or costly delay elements, and can be implemented in a wide variety of systems using simple circuit elments only.
\end{abstract}

\section{\bf INTRODUCTION}

There has been some increasing interest in recent years in the study of controlling chaotic nonlinear systems. The possibilty of obtaining periodic waveforms from a chaotic system by stabilising any of the numerous embedded Unstable Periodic Orbits (UPO's) has been the guiding control philosophy. The breakthrough in this direction is the OGY algorithm \cite{OGY}, which stabilises the UPO by applying ocassional, small, well calculated perturbations to the system parameters. A variation of this method is the Ocassional Proportional Feedback (OPF) \cite{roy}, which gives ocassional perturbation to the system parameters, proportional to the deviation of the system away from the UPO. 

	The above methods are in essence discrete in nature, and calculates the perturbation based upon local behavior of the system in a neighborhood of
the UPO. This is an advantage, as knowledge about the entire global dynamics is not necessary. However since control is exercised only ocassionaly in a small neighborhood of the UPO, the system becomes susceptible to ocassional bursts away from the UPO under moderate noise. To overcome this limitations the idea of time continous control was proposed by Pyragas \cite{pyragas}. Among the 
two algorithms suggested by Pyragas one required an external periodic signal
approximating the UPO, while the other recovered the periodic signal by Delay
Coordinate method. Though various implementations of the second method has 
been reported, the inspiration behind the present report is to replace
the delay element which is difficult and costly to obtain at some time scales.

	The present  algorithm also achieves control by perturbing the system
parameters in proportion to the error signal between the output signal to be
controlled and a periodic signal. But the periodic signal is derived from
the chaotic attractor itself, by passing the chaotic output of the system through
a band pass filter with a narrow pass band. The filter may 
be a simple LTI one, which can be implemented using resistor, capacitor and opamps only. It has been observed that though LTI filters are unable to filter out a periodic signal from a chaotic one, when connected in the above configuration, the whole system can be controlled to a periodic orbit.


\section{\bf CONTROL ALGORITHM} 

 Let us consider a dynamical system given by the model:
 \begin{equation} \label{sys}
 \begin{array}{rrr}
 \dot X&=&f(X,p)\\[2mm]
  y&=& CpX
 \end{array}
 \end{equation}
 
Where, $p$ is a parameter available for perturbation and $y$ is the output
signal to be controlled.We assume both $y$ and $p$ to be scalars. The system 
is connected in the configuration shown in Fig 1.

	The filter in the feedback path is band pass type with a narrow
pass band. A second order notch filter is found to be sufficient for effective
control. To select the pass frequency of the filter, a FFT of the output signal
is obtained, which would be typically spread spectrum, with a broad peak centred
around frequencies corresponding to the UPO's. We select the pass frequency
of filter to be within this window. A sufficiently high $Q$ value for the filter
is selected and assumed to be tunable. If the output of the filter is $y_f(t)$
the perturbation applied to the system parameter p is of the form
\begin{equation} \label{pert}
\begin{array}{rrr}
\Delta p(t)=K(y_f(t) - y(t)) 
\end{array}
\end{equation}
The value of gain $K$ is tuned to obtain stabilisation.
	
	When stabilisation is achieved the output of the filter and the system
both become periodic and close to each other; and, the perturbation $ \Delta p(t) $ becomes extremely small. Therefore, as well as in OGY and Pyragas's method small external force is used for stabilsation. Also since the pass frequency of the filter was chosen to be in the same window as that of the UPO's - the stabilised orbit lies in a small neighborhood of the UPO of the original system. The smallness of the perturbation signal depends on how close to UPO the system is stabilised and the noise level present.


	Though general validity and sufficiency conditions for the effectiveness
of the above algorithm is difficult to prove, it appears that stabilisation in the above method is achieved through additional degrees of freedom introduced
into the system by the filter in the feedback loop. The filter does not change
much the projection of the system dynamics into the original low dimensional state
space, but change only the Lyapunov exponent of the UPO. As compared to the 
delayed feedback method suggested by Pyragas \cite{pyragas}, where the system dimension increases to infinity, a finite increase in system dimension occurs
here, i.e. the finite dimensional LTI filter aproximates the delay element in some sense. 

If the state space realisation of the filter is
\begin{equation} \label{filt}
\begin{array}{rrr}
{\dot X}_f & = & A_fX_f + B_fu \\
 y_f & = & C_fX_f
\end{array}
\end{equation}

and the system equations for the chaotic dynamics are given by Eqn \ref{sys}. 
Then the state equations for the augmented system becomes

\begin{equation} \label{aug}
 \left[ \begin{array}{c} 
        \dot X \\ 
        {\dot X}_f 
         \end{array} \right]
 = \left[ \begin{array}{c} 
        f(X, p_0+K(C_fX_f-C_pX)) \\
	B_fC_pX+A_fX_f
         \end{array} \right]
\end{equation}

Stability of the UPO's can be studied by linearising the augmented system
about a fixed point and studying the singular values of the resulting Jacobian matrix.

\subsection{Stabilisation to higher periodic orbits}
It has been observed that the system could be stabilised to period-2 or
higher periodic orbits if the frequency components of these orbits are 
close enough. Stabilisation can be achieved in this case by shifting the pass frquency of the filter towards the edge of the pass window of the original system and relaxing the $Q$ value. An alternative approach could be using parallel combination of filters.

\section{\bf RESULTS FOR SOME GENERIC SYSTEMS}
The Lorentz system is used to illustrate the main results for the algorithm.
The system equations for the Lorentz system are given by
\begin{equation}
\begin{array}{rrr}
\dot x_1&=&-x_1 $ + $ x_2x_3\\[2mm]
\dot x_2&=&3(x_3 $ - $ x_2)\\[2mm]
\dot x_3&=&-x_1x_2 $ + $ rx_2 $ - $ x_3
\end{array}
\end{equation}

The state $x_2$ is selected as the output signal to be controlled and $r$ is
the system parameter available for perturbation. The nominal value of $r$ is taken as $26.0$ .

	The second order filter with transfer function \\
\[ F(s) = \frac{C\omega_0}{s^2 + \frac{\omega_0}{Q}s + \omega_0^2} \] \\
is used in the feedback loop. The FFT spectrum of the output signal is shown in Fig 2b. Accordingly the pass frequency of the filter is choosen as $ 1.2 Hz $, which lies in the pass window of the chaotic system. Stabilisation of the period-1 orbit is obtained at $Q=8.0$, $K=0.52$ and $C=0.90$. Fig 3b. and 3a. shows the stabilsed output signal and the perturbation in the system parameter respectively. Fig 3c. shows the FFT spectrum of the controlled output. 
	
	It is to be noted that the stabilised period-1 orbit is {\it almost } and not truly sinusoidal. A small harmonic component is present in the output, though topologically this can be considered an period-1 orbit. This leads to the question, whether the objective of control of chaos should be stabilisation to sinusoidal outputs, or {\it not sinusoidal} but still period-1 orbits. The answer depends on the application involved. The control algorithm suggested above offers a direct control over the harmonic content  of the output signal and can modify it according to application demand. 

	To illustrate the validity of the algorithm for other systems the results
for stabilisation of the Rossler system  to period-1 orbit is shown in Fig 4.

	 Local stability of the system can be studied by linearising the augmented state equations [\ref{aug}] about the stabilised UPO, and considering the singular values of the resulting Jacobian matrix. For the Lorentz system it is observed that the singularvalues become negative for $K \in [0.36,0.55]$ for $Q=8 $ and $f_0=1.2Hz$. The variation of singularvalues with $K$ is shown in Fig 6b.

\section{\bf CONCLUSION}	
	The above algorithm offers in a naive form a paradigm in which the problem of control of chaotic systems can also be viewed as that of synchronisation of two back to back, mutually coupled system, of which one may be a chaotic system with numerous embedded UPO's and the other may be a LTI or a simpler nonlinear system whose natural dynamics is periodic. By synchronisation of outputs of both the systems (by connecting them in feedback with a coupling gain) and tuning the simpler system such that its dynamics is close to an UPO of the chaotic system, the chaotic system can be stabilised to a periodic orbit. It is also observed that the above algorithm is more effective in controlling systems having homoclinic attractors, which points towards its similarity with non-linear PLL's. The condition on the LTI system for asymptotic synchronisation and stabilisation to UPO, is under investigation and will be reported elsewhere. There is scope for future studies in using higher order filters and nonlinear systems in the feedback path.

	In conclusion, a method of stabilising chaotic systems approximately to a UPO, by small continous time perturbations, is presented. The main advantage of the
above method lies in its easy applicabilty to a wide variety of systems. No 
complex calculations are involved and the algorithm can be implemented using low
order LTI filters only.

\section{\bf Acknowledgement}
Author would like to thank Dr.M.Vidyasagar, Director, Institute for Robotics and Intelligent Systems, for his constant encouragement and help in the course of the above work.

\newpage


\begin{figure}[htb] \label{block}
\centerline{\psfig{figure=block.eps}}
\caption{Block Diagram of the Control Algorithm}
\end{figure}

\begin{figure} [htb] \label{lc}
\centerline{\psfig{figure=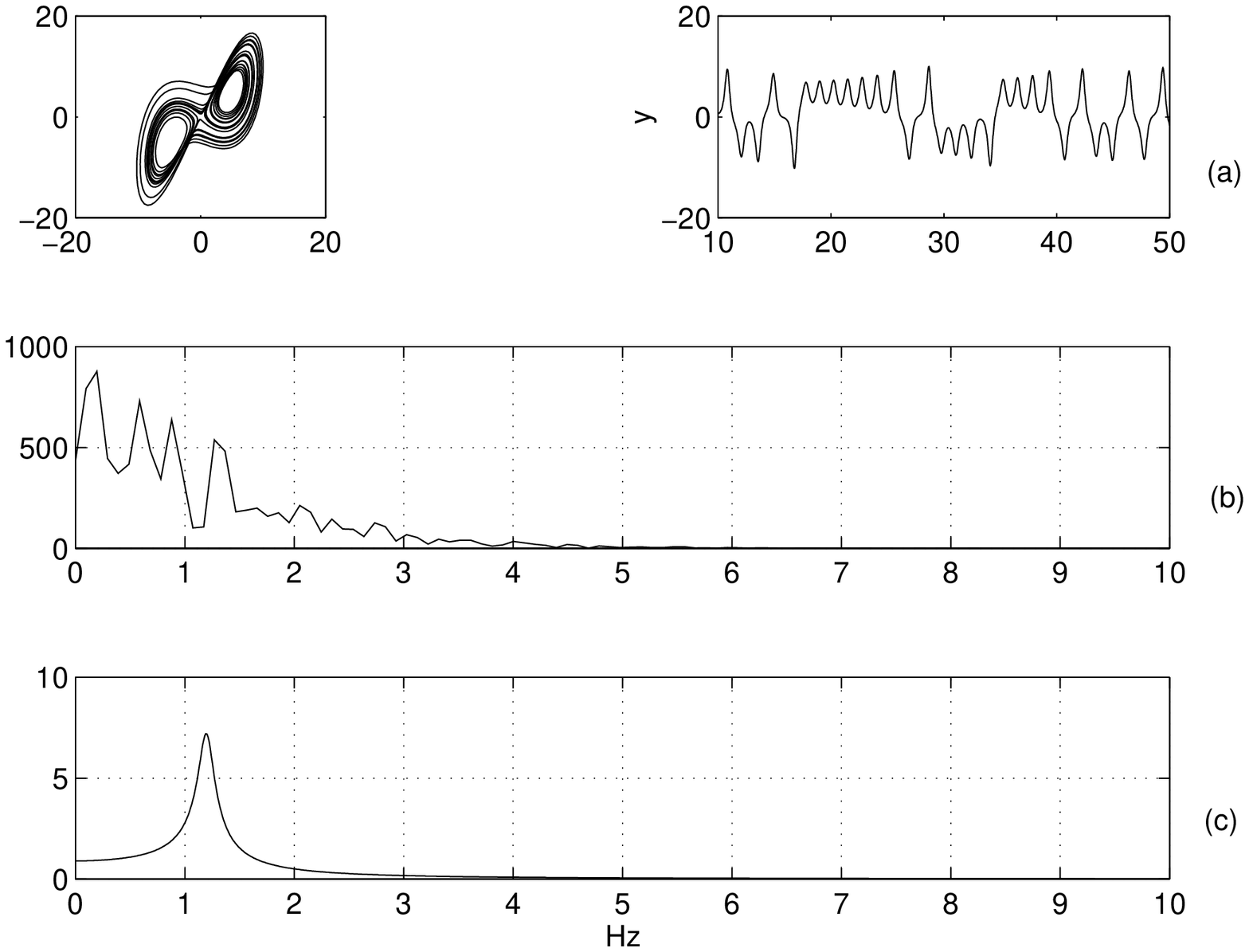}}
\caption{Dynamics of the uncontrolled Lorentz system. (a) Phase potrait and output ($x_2$) for the uncontrolled system. (b) FFT spectrum of the output over a time scale of $50s$, sampled at $50Hz$, (c) Frequency response of the filter used in feedback path.}        
\end{figure}

\begin{figure} [htb]
\centerline{\psfig{figure=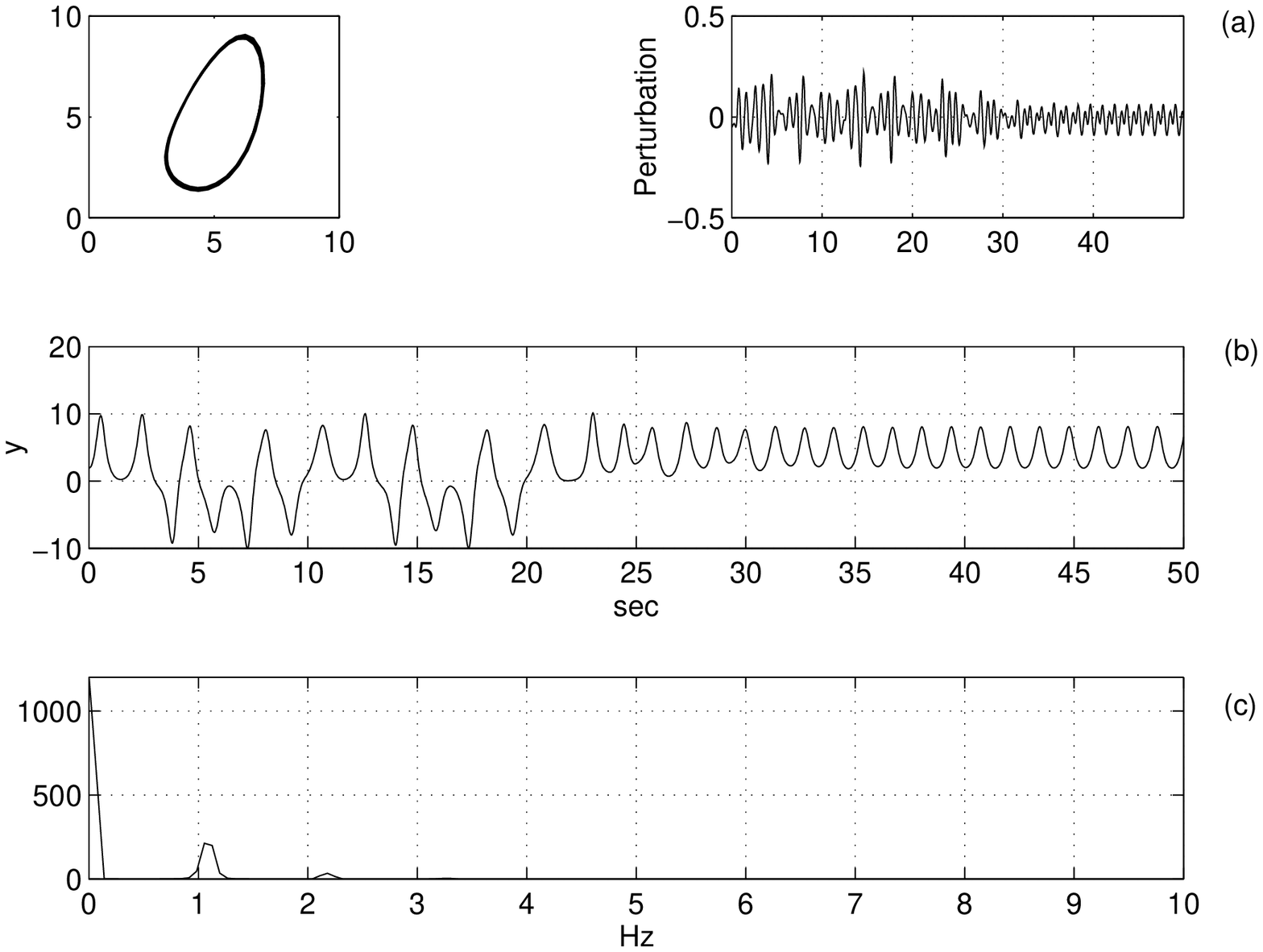}}
\caption{Controlled period-1 orbit for the Lorentz system: (a) Phase Potrait and perturbation in $r$, (note that unlike Pyragas's method where perturbation can be large during the transient period, leading to problems like multistabilty, here perturbation is small even during the transients). (b) Output signal. (c) FFT spectrum of the output after the transients are over, (the peak at dc is due to the dc bias of the output), the stabilised orbit is an {\it almost} p-1
orbit, a small higher frequency component is present along with the main frequency component at $f \approx 1.2Hz$.}
\end{figure}

\begin{figure} [htb]
\centerline{\psfig{figure=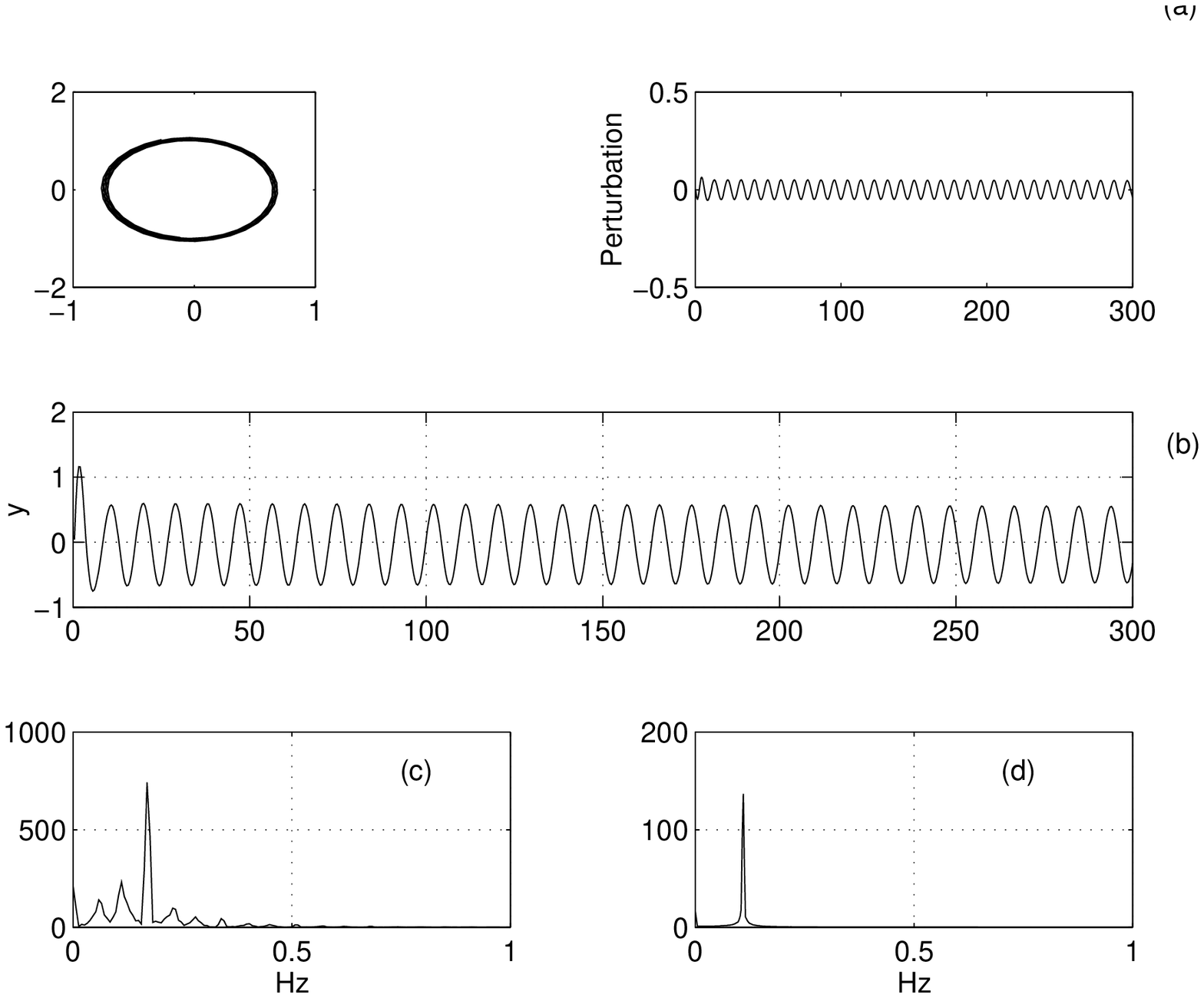}}
\caption{Controlled dynamics for the Rossler system: $ \dot{x_1}=-x_2-x_3,      \dot{x_2}=x_1+0.2x_2+\Delta p(t),  \dot{x_3}=0.2+x_3(x_1-5.7)$, perturbation $\Delta p(t)$ is determined by Eqn.2. with $y(t)=x_2(t)$ (a) Time plot of $\Delta p(t)$ (b) Controlled output. (c) FFT spectrum of the uncontrolled output. (d) FFT spectrum of the controlled output. Control is achieved at $f_0=0.12Hz, Q=1.8, C=1.0, K=0.473$. }
\end{figure}

\begin{figure} [htb]
\centerline{\psfig{figure=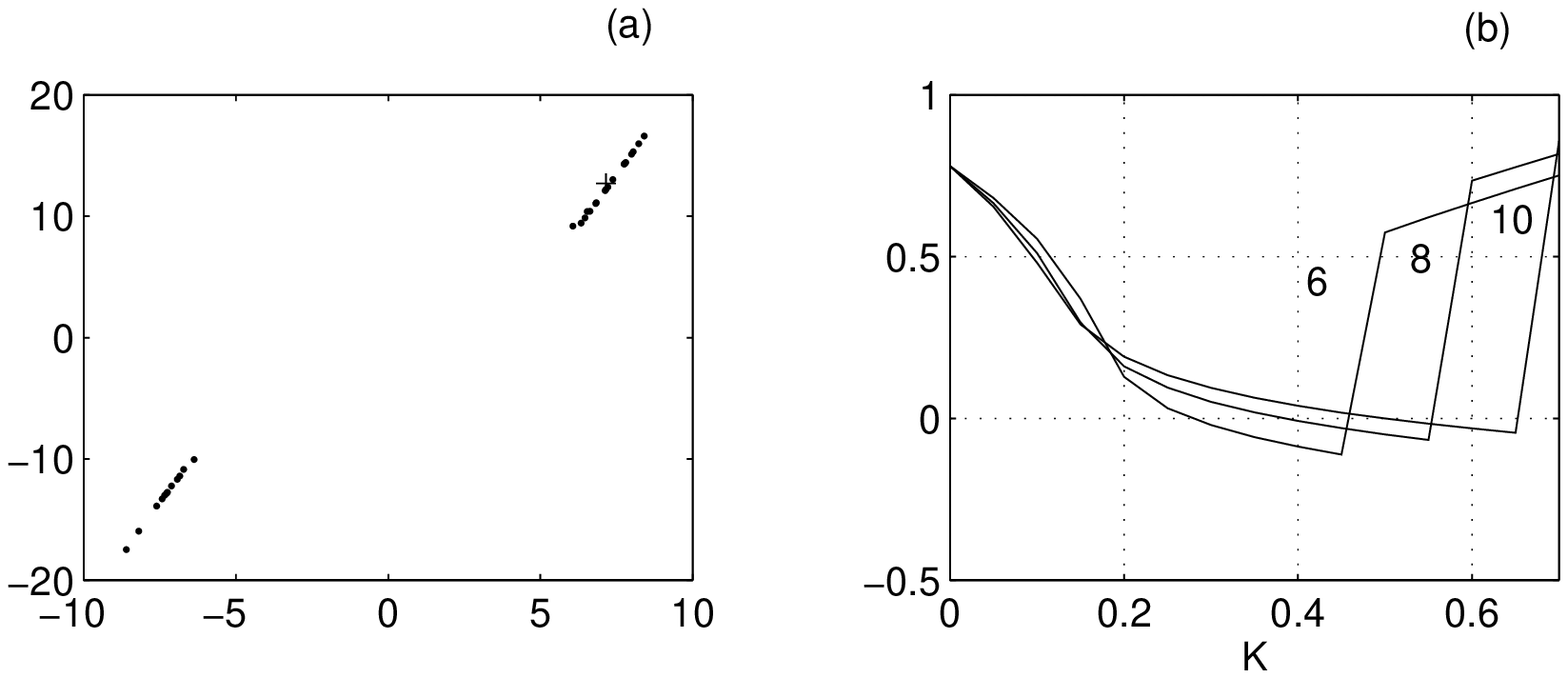}}
\caption{(a) Poincare section for the uncontrolled Lorentz system obtained at $x_1=25.0$ , '+' shows the poincare section for stabilised period-1 orbit, note that the stabilised point lies close to the fixed point of the original system (which appears denser in the figure). (b) Singular values of the Jacobian matrix for the augmented system taken about the period-1 fixed point shown in poincare section. Plots are drawn for the variation of singular values over K, for $f_0 = 1.2Hz$ and for $Q= 6,8,10.$ respectively. The value of $K$ is to be chosen in the region where the singularvalues are negative.}
\end{figure}

\end{document}